\documentstyle[prl,aps]{revtex}
\begin{document}
\draft
\title{Constrained Superconducting Membranes}
\author{Rub\'en Cordero$^{\dagger \ast}$ and 
Efra\'{\i}n Rojas$^{\dagger}$}
\address{$^{\dagger}$~Departamento~de~F\'{\i}sica,~%
Centro~de~Investigaci\'on~y~de~Estudios~Avanzados~del~IPN\\
Apdo. Postal 14-740, 07000 M\'exico D.F., MEXICO}
\address{$^{\ast}$~Departamento~de~F\'{\i}sica,~%
Escuela~Superior~de~F\'{\i}sica~y~Matem\'aticas~del~IPN\\
Edificio 9, 07738, M\'exico D.F., MEXICO}
\maketitle

\begin{abstract}
We present a geometrical canonical description for
superconducting membranes. We consider a general
action which includes a general class of superconducting 
extended objects (strings and domain walls).
The description is inspired in the ADM framework of general
relativity but, instead of the standard canonical
variables a different kind of phase space is considered.
 The Poisson algebra of the constraints and the
counting of degrees of freedom is performed. The new 
description is illustrated considering a superconducting 
domain wall on a curved background spacetime.
\end{abstract}


\section {Introduction}

Recently a number of outstanding problems in physics 
are currently being solved using geometrical techniques
involving the machinery of differential geometry. In this 
way, superconducting strings moving on a
curved background spacetime have also been the subject of
intense research because it believed that they appear in the 
early Universe as topological defects \cite {Vilenkin}.  
However, domain walls can exhibit a superconducting
character, too. Superconducting domain walls appear in 
supersymmetry and grand unified theories \cite{Lazarides}. 
In adittion, domain walls transform in superconducting 
membranes in a similar way like in Witten's superconducting 
string \cite{Peter}.

Among the most important aspects around extended objects is the
question of its dynamical behavior. For example, in the case of 
circular strings, the dynamics shows us when the system
exhibit string collapse from radial configuration or not.
A convenient way to deal with the last question  
lies in the Hamiltonian formulation since it cast out a 
form of the effective potential for the relativistic membrane. 
It could help us to separate the equilibrium configurations 
from the unequilibrium (collapsing) ones \cite{Larsen}.
In the Hamiltonian context, superconducting strings evolving 
in a curved background spacetime have been explored in 
\cite{Larsen,Carter,Boiss}

The purpose of this article is to study the Hamiltonian formulation 
for superconducting membranes in a geometric language without 
specification of any special membrane configuration and 
particular background geometry. We get a truly 
canonical formulation for this 
system which involves the full machinery of an Hamiltonian 
formulation, i.e., the specification of an appropiate set of 
phase space variables, of a Poisson bracket structure and a 
Hamiltonian function. This description possess 
a rich geometric content. To our knowledge this geometrical 
canonical analysis for superconducting membranes has not 
been considered  before. Another goal of the work, and important
application, is the preparation of the classical theory in a 
suitable form for its canonical quantization. In fact, the 
investigation of the structure of classical dynamics of totally 
constrained systems will shed a good light on how to find a
consistent
quantum formulation of them. Quantum mechanics aspects in 
superconducting strings have been reported in
\cite{Davidson,Axenides}.
The paper is organized as follows: In Sec. II we develop
the mathematical issues needed for the ADM fashion in order 
to achieve the Hamiltonian formulation. In Sec. III we 
perform the Hamiltonian formulation of our system. The 
variation of the phase space constraints and their physical
meaning is done in Sec. IV. The constraint algebra is 
presented in Sec. V. Finally, in Sec. VI, we 
treat an specific example that illustrates our results.  

\section {Mathematical Tools}

We consider a relativistic membrane of dimension
$d$ (for strings $d=1$ and for domain walls $d=2$). Its
worldsheet, 
$m$, of dimension $d+1$ is an oriented timelike surface 
embedded in an arbitrary fixed $4$-dimensional background 
spacetime $\{ M , g_{\mu\nu} \}$. Following the ADM treatment 
of canonical general relativity \cite{ADM}, we assume that $m$
has the 
topology $\Sigma \times R$ such that we can consider a 
foliation of the worldsheet $\{ m , \gamma_{ab} \}$  into spacelike 
hypersurfaces of dimension $d$, $\Sigma_t$, defined by the 
constant value of a certain scalar function $t$. Each leaf 
of the foliation represents the system at an instant of 
time, and each one is diffeomorphic to each other. The
hypersurface
$\Sigma_t$ may be described by the embedding
of $\Sigma_t$ into $m$
\begin{equation}
\xi^a = X^a (u^A)\,,
\label{eq:1}
\end{equation}
where $u^A$ are local coordinates on $\Sigma_t$ and $\xi^a$
are local coordinates on $m$ \,$(a,b=0,...,d+1)$ and
($A,B=1,\cdots, d$).
Locally, one can think of a decomposition
of the worldsheet embedding functions $\chi^\mu$
into a timelike coordinate $t$, and spacelike
coordinates $X^a$ where $\chi^\mu$ are the embedding functions
of $m$ on $M$. 

Alternatively, it is useful to
consider the direct embedding, via map composition,
of $\Sigma_t$ into the background spacetime
$\{ M , g_{\mu\nu} \}$,
\begin{equation}
x^\mu = X^\mu (u^A) = \chi^\mu(\xi^a (u^A))\,,
\label{embedd}
\end{equation}
where $e^\mu {}_a$ are the tangent vectors to $m$ associated
with the embedding $\chi^\mu$. The $d$ tangent vectors to 
$\Sigma_t$ are defined with
\begin{equation}
\epsilon_A = \epsilon^a{}_A e^\mu{}_a 
\partial_\mu \,, 
\label{eq:2b} 
\end{equation}
so that the positive-definite metric induced on $\Sigma_t$
is
\begin{equation}
h_{AB}= \gamma_{ab}\epsilon^a{}_A \epsilon^b{}_B 
= g_{\mu\nu}\epsilon^\mu{}_A \epsilon^\nu{}_B \,\,,
\label{eq:3b}
\end{equation}
where $ \epsilon^a{}_A $ denotes the tangent vectors to
$\Sigma_t$
associated with the embedding~(\ref{eq:1}).
The unit, future-directed, timelike normal to $\Sigma_t$,
$\eta^a$,
is defined, up to a sign, with 
\begin{equation}
\gamma_{ab}\eta^a \epsilon^b{}_A  = 0\,, \;\;\;\;
\gamma_{ab}\eta^a \eta^b  = - 1\,.
\end{equation}
In order to describe the evolution of the
leaves of the foliation, we define
a worldsheet time vector field with
\begin{equation}
t^a = N \eta^a + N^A \epsilon^a{}_A\,,
\end{equation}
where $N$ is called the lapse function, and
$N^A$ the shift vector.
The deformation vector of a spacelike hypersurface
in spacetime
is simply the push-forward of $t^a$, $\dot{X}^\mu := t^a
e^\mu{}_a$.
A basic step in the canonical analysis of the most 
important physical theories is the decomposition of 
the several geometric quantities involved in the 
theory in the normal to and tangential 
parts to an embedded hypersurface $\Sigma_t$. Thus, the
worldsheet metric, $\gamma_{ab}$, can be decomposed
in the standard  ADM fashion as
\begin{equation}
\gamma_{ab} = 
\left(
\begin{array}{ll}
 - N^2 + N^A N_A & N_A \\
\,\,\,\,\,\,\,\,\,\,\,\,\,\, N_A & h_{AB} 
 \end{array}
\right)\,.
\label{gamma}
\end{equation}
and for the inverse
\begin{equation}
\gamma^{ab} = 
{1 \over N^2} \left(
\begin{array}{ll}
 - 1  & \hspace{1.5cm}N^A  \\
 N^A   & (h^{AB} N^2 - N^A N^B )  
\end{array}
\right)\,.
\end{equation}
Note that it follows from the expression~(\ref{gamma}) that
\begin{equation}
\mbox{det} (\gamma_{ab}) = - N^2 \,\mbox{det} (h_{AB})\,.
\end{equation}
Furthermore, we denote with ${\cal D}_A$ the (torsionless)
covariant derivative compatible with $h_{AB}$. 
For a more general treatment about this kind of geometrical
decomposition in extended objects see the reference 
\cite{Rojas}.

\section {Hamiltonian Formalism}

We consider superconducting membranes described by 
the generic effective action
\begin{equation}
S = \int_m \sqrt{- \gamma}\,{\cal L}(\omega)
\label{act}
\end{equation}
where ${\cal L}(\omega)$ is
a function depending of the internal and external fields 
acting on the membrane through $\omega = \gamma^{ab}\,
(\nabla_a \phi + A_\mu e^\mu {}_a ) (\nabla_b \phi + 
A_\mu e^\mu {}_b ) $\,,
which is the worldsheet projection  of the gauge covariant derivative of a worldsheet scalar field $\phi$ and 
$A_\mu$ is the background electromagnetic potential. 
Note that it has absorbed the worldsheet 
differential $d^{d+1}\xi$ into the integral sign. Similarly,
we will do the same for integrals over $\Sigma_t$.
The ADM decomposition of the action ~(\ref{act}) implies the 
Lagrangian 
\begin{equation}
L[X,\dot{X};\phi, \dot{\phi} ] = \int_{\Sigma_t} N\,\sqrt{h}
\,{\cal L}(\omega)\,.
\end{equation}
From now on, we should understand ${\cal L}$ as in the 
ADM fashion. 
Besides, the ADM decomposition of $\omega$ is given by
\begin{equation}
\omega = - \frac{1}{N^2}\,[ {\cal L}_t \phi - N^A 
\tilde{{\cal D}}_A \phi + (t\cdot A) ]^2 + 
\tilde{\omega}\,\,, 
\label{omega}
\end{equation}
where we have defined the $\Sigma_t $-projection of 
the gauge covariant derivative of the worldsheet 
scalar field $\phi$, i.e.,
$\tilde{{\cal D}}_B \phi := {\cal D}_B \phi + A_B $ 
with $A_B = A_\mu \epsilon^\mu {}_B$ and, 
\, ${\cal L}_t \phi$
denotes the Lie derivative of $\phi$ along the 
deformation vector field
$t^\mu$ and $\tilde{\omega}=  h^{AB} \tilde{{\cal D}}_A 
\phi \tilde{{\cal D}}_B \phi$ is the $\Sigma_t$-projection 
of $\omega$ . 

The momenta are defined as 
\begin{eqnarray}
\pi &=& \frac{\delta L}{\delta {\cal L}_t \phi}  
 =- 2 \sqrt{h} (\tilde{\omega} - \omega )^{1/2}
\frac{d {\cal L}}{d \omega} \\
P_\mu &=& \frac{\delta L}{\delta \dot{X}^{\mu}} 
= [ - \sqrt{h} {\cal L}(\omega) + (\tilde{\omega} - 
\omega )^{1/2}\,\pi ]\,\eta_\mu - \pi \,
\tilde{{\cal D}}_A \phi \,\, 
\epsilon_\mu {}^A  + \,\,\,\pi  A_\mu \,.
\end{eqnarray}
The phase space for our superconducting membrane is 
naturally associated with the geometry of $\Sigma_t$, 
$\Gamma:= \{ X^\mu,P_\mu; \phi , \pi \}$. 
In order to get more simplicity we define the kinetic 
momentum $\Upsilon_\mu$ as follows, 
$\Upsilon_\mu := P_\mu - \pi A_\mu $. The action~(\ref{act}) 
is invariant under reparameterization of the worldsheet 
so, we should expect to have phase space constraints that
generate this gauge freedom. Indeed, from the definition 
of the momenta we have the constraints 
\begin{eqnarray}
C_0 &=& g^{\mu \, \nu}\Upsilon_\mu \Upsilon_\nu + 
h\,\left( {\cal L}(\omega ) + \frac{\pi ^2}{2\,h\,
(d{\cal L}/d\omega )}
\right)^2  -  \pi^2 \,\left( \omega + \frac{\pi^2}
{4\,h\,(d{\cal L}/d\omega)^2}
\right) 
\label{C0} \\
C_A&=& \Upsilon_\mu \,\epsilon^\mu {}_A + \pi\,
\tilde{{\cal D}}_A \phi \,.
\label{C1} 
\end{eqnarray}
The former, or scalar constraint, is the generalization
to superconducting membranes of the scalar constraint 
for a parametrized relativistic particle in external electromagnetic field. The latter, or vector constraint, 
is universal for all reparametrizations invariant actions 
of first order in derivatives of the embedding functions. 
In fact,~(\ref{C0}) is that depends from a particular 
form of the Lagrangian. It should be noted that for a superconducting cosmic string in a stationary background 
the constraint~(\ref{C0}) reduce to eq. (24) in 
\cite{Larsen}, hence our result is more general. Observe 
that the constraints~(\ref{C0})
and~(\ref{C1}) are valid for higher dimensions, $d > 2$, 
but unfortunately, nowadays there is not an effective 
theory describing such higher extended objects, at the 
most, strings and domain walls.

The constraints perform a double duty. Their first job
is to restrict the possible values of the phase space
variables, i.e, they tell us that there is a redundancy 
in the characterization of a field configuration in 
terms of phase space points. The second job of the 
constraints is that, before they are set to zero, they 
generate dynamics. They are the generators of the 
canonical evolution of the system.
 
In order to obtain the geometric information encoded 
in the constraints~(\ref{C0}) and~(\ref{C1}), we 
recast them as functionals on $\Gamma$ (called 
{\it constraint functions}).
To do this, we smear them with test fields $ \lambda $ 
and $\lambda^A$, defining a scalar constraint function:
\begin{equation}
{\cal S}_{\lambda} [ X,\phi ,P,\pi ]= \int_{\Sigma_t} 
\lambda \left[ g^{\mu \, \nu}\Upsilon_\mu \Upsilon_\nu + 
h\,\left( {\cal L}(\omega ) + \frac{\pi ^2}{2\,h\,
(d{\cal L}/d\omega )} \right)^2  
- \,  \pi^2 \,\left( \omega + 
\frac{\pi^2}{4\,h\,(d{\cal L}/d\omega)^2} \right) \right] \,,
\label{esc}
\end{equation}
and a vector constraint function
\begin{equation}
{\cal V}_{\vec{\lambda}} 
[ X,\phi,P,\pi ] = \int_{\Sigma_t} 
\lambda^A  \left( \Upsilon_\mu \,\epsilon^\mu {}_A + \pi\,
\tilde{{\cal D}}_A \phi \right)\,.
\label{vec}
\end{equation}
The smearing field $\lambda$ must be a scalar density 
of weight minus one since the scalar constraint function 
should be well defined. Using Hamilton equations, we can 
show how $\{ \lambda , \lambda^A \}$ are related
to the lapse function and the shift vector.

Note that our Hamiltonian vanishes, ${\cal H}=0$. This 
was to be expected from reparameterization invariance.
However, according to the standard Dirac
treatment of constrained systems,
the Hamiltonian vanishes only weakly \cite{Dirac}. 
It is a linear combination of the constraints,\,\,
$H [X,\phi,P,\pi] = \int_{\Sigma_t}  ( \lambda C_0 
+ \lambda^A C_A )$\,.

\section{Constraints}

For purposes of reach a true canonical description
of the superconducting membranes, in the following 
subsections we discuss the variation of the phase 
space constraints. The variation of the
action with respect to the phase space variables
contains the variations of each constraint function 
which contribute a surface term. They are neglected 
for lack of simplicity.

\subsection{Vector Constraint}

The Hamiltonians vectors fields generated 
by~(\ref{esc}) and~(\ref{vec}) should correspond 
to evolution along the vectors fields $N\eta^{\mu}$ 
and $N^A$, respectively.
To evaluate the Poisson brackets of the constraints and to 
determine the motions they generate on phase space we
compute, first, the infinitesimal canonical tranformations
generated by the vector constraint ${\cal V}_{\vec{\lambda}} 
$. Modulo some boundary conditions we get the functional
derivatives
\begin{eqnarray}
\frac{\delta {\cal V}_{\vec{\lambda}}}{\delta X^\mu}&=&
- {\cal D}_A (\lambda^A\,p_\mu)= - {\cal L}_{\vec{\lambda}}
 \,p_\mu 
\hspace{1.5cm}
\frac{\delta {\cal V}_{\vec{\lambda}}}{\delta \phi}=
- {\cal D}_A (\lambda^A\,\pi )\,\,= - {\cal L}_
{\vec{\lambda}}\, \pi \\
\frac{\delta {\cal V}_{\vec{\lambda}}}{\delta p_\mu}&=&
\lambda^A \,\epsilon^\mu {}_A  \,\,\,\,\,\,\,\,\,\,\,\,\,= 
{\cal L}_{\vec{\lambda}} \,X^\mu 
\hspace{1.8cm}
\frac{\delta {\cal V}_{\vec{\lambda}}}{\delta \pi}=
\lambda^A \,{\cal D}_A \phi \,\,\,\,\,\,\,\,\,\,\,= 
{\cal L}_{\vec{\lambda}} \,\phi \,,
\end{eqnarray}
where ${\cal L}_{\vec{\lambda}}$ denotes the Lie derivative 
along the vector field $\lambda^A$. The Hamiltonian vector 
field generated by ${\cal V}_{\vec{\lambda}}$ is 
\begin{equation}
W_{{\cal V}_{\vec{\lambda}} } = \int_{\Sigma_t} 
\left[ ({\cal L}_{\bf \vec{\lambda}} \,X^\mu ) 
{ \delta \over \delta X^\mu }
+ ({\cal L}_{\bf \vec{\lambda}} \,\phi ) 
{ \delta \over \delta \phi }
+ ( {\cal L}_{\bf \vec{\lambda}}\, P_\mu ) 
{ \delta \over \delta P_\mu }
+ ( {\cal L}_{\bf \vec{\lambda}} \,\pi ) 
{ \delta \over \delta \pi }
\right] \,.
\end{equation}
This is consistent with the geometrical interpretation 
of the vector constraint ${\cal V}_{\vec{\lambda}} $ 
as the phase space generator of diffeomorphisms on 
the surface $\Sigma_t$. Acting $W_{{\cal V}_
{\vec{\lambda}} }$ on each one 
of the phase space variables we get  
\begin{eqnarray}
X^\mu &\to & X^\mu + \epsilon \,
{\cal L}_{\bf \vec{\lambda}}\, X^\mu \hspace{2cm}
\phi \to \phi + \epsilon \,{\cal L}_
{\bf \vec{\lambda}}\, \phi   \nonumber \\
P_\mu &\to & P_\mu + \epsilon \,
{\cal L}_{\bf \vec{\lambda}}\, P_\mu \hspace{2.3cm}
\pi \to \pi + \epsilon \,{\cal L}_{\bf \vec{\lambda}}\, 
\pi  \,, \nonumber 
\end{eqnarray}
which is the motion on $\Gamma$ generated by 
${\cal V}_{\vec {\lambda}}$. Here $\epsilon $ is an 
infinitesimal parameter.

\subsection{Scalar Constraint}

We turn now to the scalar constraint. 
The computation of its Hamiltonian 
vector field is more complicated due 
to the several variations involved. In a similar 
way to vector constraint, from the equation~(\ref{esc}), 
after tedious algebra we get the functional derivatives
\begin{eqnarray}
\frac{\delta {\cal S}_\lambda}{\delta X^\mu}&=&
- 2 \,\lambda \, (p^\nu - \pi A^\nu )\,\pi \,A_{\nu \,, 
\,\mu} + 2\,\lambda \,h\,{\cal L}^+ (\omega) {\cal L}
(\omega)\,K^I m_\mu {}^J \,\eta_{IJ}   \nonumber \\
&&- \,\,{\cal D}_A(2\,\lambda \,h\,{\cal L}^+ (\omega) 
{\cal L}(\omega))\,
\epsilon_\mu {}^A - 4\,\lambda \,h\,{\cal L}(\omega)\,
\frac{d {\cal L}}{d \omega} \tilde{{\cal D}}_A \phi \,\,
\tilde{{\cal D}}_B
\phi \,K^{AB\,I}m_\mu {}^J \,\eta_{IJ} \nonumber \\
&&+ \,\,{\cal D}^A (4\,\lambda \,h\,{\cal L}(\omega)\,
\frac{d {\cal L}}{d \omega} \tilde{{\cal D}}_A \phi \,
\tilde{{\cal D}}_B
\phi ) \,\epsilon_\mu {}^B - {\cal D}_A (4\,\lambda \,h\,
{\cal L}^+
(\omega)\,\frac{d {\cal L}}{d \omega}\,h^{AB}\, 
\tilde{{\cal D}}_B \phi
\,A_\mu ) \nonumber \\
&&+ \,\,4\,\lambda \,h\,{\cal L}(\omega)\,
\frac{d {\cal L}}
{d \omega}\,h^{AB}\, \tilde{{\cal D}}_A \phi \,\,
\epsilon^\nu {}_B\,
\,A_{\nu \,,\,\mu}  - {\cal D}_A (2\,\lambda \,
h^{AB}\,(p_\mu
- \pi A_\mu ) \,\pi \,\tilde{{\cal D}}_B \phi ) \\
\frac{\delta {\cal S}_\lambda}{\delta \phi}&=&
- {\cal D}_A (4\,\lambda \,h\,{\cal L}^+ (\omega)\,
\frac{d {\cal L}}{d \omega}\,h^{AB}\, \tilde{{\cal D}}_B 
\phi)    \\
\frac{\delta {\cal S}_\lambda}{\delta p_\mu}&=&
2\,\lambda\,(p^\mu - \pi A^\mu ) + 2\,
\lambda\,h^{AB}\,\pi \,
\tilde{{\cal D}}_A \phi \,\,\epsilon^\mu {}_B \\
\frac{\delta {\cal S}_\lambda}{\delta \pi}&=&
- 2\,\lambda\,(p^\mu - \pi A^\mu )\,A_\mu + \lambda \,
\frac{{\cal L}^+ (\omega)\,\pi }{d {\cal L} /d \omega} 
- 2\,\lambda \,
h^{AB}\, \pi \,\tilde{{\cal D}}_A \phi \,A_B \,\,,
\end{eqnarray}
where we have defined the quantity ${\cal L}^+ (\omega)
:= {\cal L}(\omega) + \frac{\pi ^2}{2\,h\,
d {\cal L}/ d \omega} $\,\,, $K_{AB} ^I$ denotes 
the extrinsic curvature of $\Sigma_t$ associated with 
the embedding~(\ref{embedd}), $m^{\mu \,I}= 
\{ \eta^\mu , n^{\mu \,i} \}$ 
is the complete orthonormal basis associated 
with~(\ref{embedd}) and $\eta_{IJ}$ is the Minkowski 
metric with signature $(- , + ,...)$,
$I=0,i$ \cite{Defoedges}. The corresponding Hamiltonian 
vector field is given by
\begin{equation}
W_{{\cal S}_\lambda}= \int_{\Sigma_t} \frac{\delta 
{\cal S}_\lambda}
{\delta p_\mu}\,\frac{\delta }{\delta X^\mu} + 
\frac{\delta {\cal S}_\lambda}{\delta \pi}\,
\frac{\delta }{\delta \phi} - \frac{\delta {\cal S}_\lambda}{\delta X^\mu}\,
\frac{\delta }{\delta p_\mu} - \frac{\delta 
{\cal S}_\lambda}{\delta \phi}\,\frac{\delta }
{\delta \pi}\,\,\,,
\end{equation}
which is the generator of time evolution on the 
worldsheet $m$. The motion on $\Gamma$ generated by 
${\cal S}_\lambda$ is obtained acting $W_{{\cal S}_\lambda}$ 
on each one of the phase space variables. In fact, 
this constraint is the hard part in the description. It 
generates diffeomorphisms out of the spatial
hypersurface which can be considered as diffeomorphisms 
normal to $\Sigma_t$. For a spatial observer this is 
dynamics. Furthermore, it is instructive to make a naive 
analysis of the scalar constraint. There is a kinetic 
term of the form $g^{\mu \nu} \Upsilon_\mu \Upsilon_\nu $. 
The potential term corresponds to an effective 
potential for the system which could help us to separate the
stable configurations from the unstables ones. 
For a circular cosmic string a effective potential 
was obtained by Larsen \cite{Larsen}.
In Sec. VI we will treat the case of a charged 
spherically symmetric membrane.

\section{Constraint Algebra}

If one is interested in the canonical quantization program
we compute the algebra of constraints since that is an
important point in the promotion of phase space variables to
operators \cite{Dirac}. Furthermore, at classical level,
constraint algebra show us how the initial value equations 
are preserved in time in the canonical language by the 
closure of the Poisson algebra of the
constraints. The Poisson bracket (PB) between any two 
functionals $f$ and $g$ of $\Gamma$ will be given by
\begin{equation}
\{ f, g \}= \int_{\Sigma_t} \frac{\delta f}{\delta P_\mu}
\frac{\delta g}{\delta X^\mu} + \frac{\delta f}{\delta \pi}
\frac{\delta g}{\delta \phi} - \frac{\delta f}{\delta X^\mu}
\frac{\delta g}{\delta P_\mu} - \frac{\delta f}{\delta \phi}
\frac{\delta g}{\delta \pi}\,\,.
\label{PB}
\end{equation}
The Dirac algebra is given by
\begin{eqnarray}
\{ {\cal V}_{\vec{\lambda}} ,{\cal V}_{\vec{\lambda'}} \} &=&
- {\cal V}_{[\vec{\lambda} ,\vec{\lambda']}  } 
\label{parent1} \\
\{ {\cal V}_{\vec{\lambda}} ,{\cal S}_{\lambda} \}&=&
- {\cal S}_{{\cal L}_{\vec{\lambda}(2\lambda)}}  
\label{parent2} \\
\{{\cal S}_\lambda ,{\cal S}_{\lambda ^{'}}\} &=&
- {\cal V}_{\vec{\lambda} ^*} \,\,,
\label{parent3}
\end{eqnarray}
where
\begin{equation}
\lambda^{*\,A}= 4h\left( {\cal L}^+ (\omega){\cal
 L}(\omega)h^{AB}
- 2{\cal L}^- (\omega)\frac{d {\cal L}}
{d \omega}\,h^{AC}\,h^{BD}\,
\tilde{{\cal D}}_C \phi \, \tilde{{\cal D}}_D 
\phi \right) \,(\lambda \,{\cal D}_B \lambda' - 
\lambda'\,{\cal D}_B \lambda )\,\,,
\end{equation}
and we have defined the quantity $ {\cal L}^- (\omega)
 := {\cal L}(\omega) 
- \frac{\pi ^2}{2\,h\,d {\cal L}/ d \omega} $.
Equation~(\ref{parent1}) 
is the algebra of spatial diffeomorphisms generated by 
${\cal V}_{\vec{\lambda}}$ \,, and it is isomorphic to 
the Lie algebra of infinitesimal spatial diffeomorphisms.
Furthermore, this algebra
is the same one would expect in any theory with gauge 
invariance. The PB~(\ref{parent2}) shows how 
${\cal S}_\lambda$ transforms under spatial 
diffeomorphisms. Finally, the crucial 
PB~(\ref{parent3}) means that two infinitesimal normal deformations on $\Sigma_t$\,, performed
in an arbitrary order, end on the same final 
hypersurface but not on the same point on that 
hypersurface. Hence, the algebra 
closes and it is first-class in the Dirac terminology \cite{Dirac}. 
Remarks are in order: {\it i)} 
Note that the right-hand side of~(\ref{parent3}) involves 
{\it structure functions} which is a source of problems in 
any attempt to use the Dirac algebra in the canonical 
quantisation program because the PB 
algebra will go over to a commutator algebra which is 
not a Lie algebra. {\it ii)} Since Hamiltonian is a 
linear combination of the constraints themselves, the 
constraints are preserved in time. {\it iii)} According 
to \cite{Henneaux}, the explicit counting of degrees of 
freedom goes as follows: $ 2 \times 
({\mbox{number of physical degrees of freedom}}) = 
({\mbox{total number of canonical variables}}) - 2 \times ({\mbox{number of first-class constraints}})$\,. Hence, 
there are $N - d $ physical degrees of freedom. {\it iv)} 
The Hamilton equations of the system are computed from the functional derivatives listed before. Their form is 
large and unaesthetic and we only mention that they are equivalent to the conservation of current density, i.e.,
$\nabla_a J^a =0$, and to the equation of motion 
of the system, namely, $T^{ab}K_{ab} ^i = F_a {}^i J^a$, 
\cite{Carter,Cordero}.

\section{Example}

In order to illustrate the results reached before, we 
now analyse the case of a charged spherically symmetric 
membrane ($d=2$) described by the action~(\ref{act}), 
which evolves in a general spherically 
symmetric static background spacetime ($N=4$)
\begin{equation}
ds^2 = -A(r)\,dt^2 + B(r)\,dr^2 + C(r)\,d\Omega^2
\end{equation}
where the functions $A,B$ and $C$ depend on the 
particular ambient spacetime. 
We take as {\it ansatz} for our membrane the choice $\phi = 
\phi(t)$ and $A_\mu = (A_0 (r),0,0,0)$. The physical 
consequence due to last is that membrane is charged 
only and no currents exist there. 
According to the embedding of $m$ into $M$, $\chi^\mu 
(t,\theta, \varphi ) = (t,r(t),\theta ,\varphi)$ the 
induced metric on the worldsheet is
\begin{equation}
\gamma_{ab} = 
\left(
\begin{array}{lll}
- A + B\,\dot{r}^2 & 0 & \,\,\,\,\,\,\,\,0 \\
\,\,\,\,\,\,\,\,\,\,\,\,\,0& C & \,\,\,\,\,\,\,\,0 \\
\,\,\,\,\,\,\,\,\,\,\,\,\,0 & 0& C\,{\mbox{sen}}^2 \theta 
\end{array}
\right) \,\,.
\end{equation}
It follows that $\gamma = (- A + B\,\dot{r}^2 )\,C^2 
\,\sin^2 \theta$. Now, the embedding of $\Sigma_t$ 
in $M$, $X^\mu (\theta,\varphi )= (t_0 ,r_0 ,\theta 
,\varphi)$, bear the induced metric
\begin{equation}
h_{AB} = 
\left(
\begin{array}{ll}
C_0 & \,\,\,\,\,\,\,\,\,0  \\
 \,0& C_0 \,{\mbox{sen}}^2 \theta 
\end{array}
\right) \,\,,
\label{superhAB}
\end{equation}
where $C_0 = C(r_0 )$, hopefully that $C_0$ in this 
section not create confusion with the constraint~(\ref{C0}). 
Observe that $h=C_0 ^2 \,\sin^2 \theta$. 
From~(\ref{omega}) we find 
\begin{equation}
\omega = \frac{(\dot{\phi} + A_0)^2}{- A + B\,
\dot{r}^2 } \,\,.
\label{omega1}
\end{equation}
According to the continuity equation $\sqrt{-\gamma}\,
\nabla_a J^a = \partial_a ( \sqrt{-\gamma}\,2\,
\frac{d{\cal L}}{d \omega}\,\gamma^{ab}\,
\tilde{\nabla}_b \phi)=0$ we can rewrite
the equation~(\ref{omega1}) as
\begin{equation}
\omega = - \frac{W^2}{4 (d{\cal L}/d\omega)^2 \,C^2}\,,
\end{equation}
where $W$ is an integration constant. 

The full information developed leads to an expression 
for the effective potential 
\begin{eqnarray}
V_{{\mbox{eff}}}&=& C^2 \,\sin ^2 \theta \,
({\cal L}^+ (\omega))^2 
\nonumber \\
&=&C^2 \,\sin ^2 \theta \,\left( {\cal L}(\omega) + 
\frac{W^2}{2 d{\cal L}/d\omega \,C^2} \right) ^2\,,
\label{potefect}
\end{eqnarray}
which is of the separable form of the kinetic term 
in~(\ref{C0}). To see this, expanding the kinetic 
part in the scalar constraint~(\ref{C0}) and adding 
to~(\ref{potefect}) we obtain a first 
integral of motion of the system
\begin{equation}
\frac{B}{A}\,\dot{r}^2 = 1 - \frac{C^2}{(E + W\,A_0)^2} 
\,({\cal L}^+ (\omega))^2 \,,
\end{equation}
where $E$ denotes the energy of the system. It is 
worthy to mention that until now the Lagrangian 
${\cal L}(\omega)$ as been treated as an arbitrary 
function of $\omega$. The several cases of the ambient
spacetime and models describing the superconducting 
extended objects is an easy task. For instance, in 
flat background spacetime we can obtain an equilibrium 
configuration with the Witten model, \,${\cal L}
(\omega)= 1 + \omega /2$. This kind of model has been 
used in an equivalent formulation in terms of a gauge 
field over the worldsheet in \cite{Guendelman1}. 
Another interesting systems have been 
considered in the literature, for instance, the case of 
charged current-carrying circular string on Kerr 
background spacetime and flat spacetime \cite{Carter,Carter1}.

\vspace{0.4cm}

\section{Conclusions}

In this work we have achieved a canonical analysis for 
superconducting membranes in a geometrical way. We find 
that the Dirac algebra is closed and therefore the 
constraints are of first-class. The canonical quantization 
for the system is performed by considering the phase space 
variables as operators and by replacing the PB by 
conmutators, where physical states will satisfy the 
condition $\hat{H}|\Psi> =0$. Furthermore, our Hamiltonian 
constraint~(\ref{C0}) is a general expression obtained in 
a geometrical way without hide any information. We 
gave as example, a charged bubble embedded in an arbitrary 
spherically symmetric static background spacetime. The separability 
of the presented system is obtained thanks 
to the symmetry, but not the integrability. In the case 
of a black hole as background, one must be careful with 
$r$ bubble coordinate with respect to $r_+$ (event 
horizont), so as to $r > r_+$. For superconducting strings 
exist a dual formalism in terms of a scalar gauge 
independent function and its canonical formulation 
can be easily extended in a similar way. However, when 
we deal with the superconducting wall new structure 
apeared due to the presence of a vector field over the worldsheet. 
The study of this subject is under current investigation.

\acknowledgments
We thank X. Martin, R. Capovilla, J. Guven and
E. Ay\'on for fruitful discussions. We are also 
grateful to H. Garc\'{\i}a-Compe\'an for reading 
the manuscript and useful comments. This work was 
partially supported by CONACyT and SNI M\'exico.

\end{document}